\documentclass[twocolumn,aps,prl,a4paper]{revtex4-1}

\usepackage{graphicx}

\usepackage{fancyhdr}
\usepackage{pslatex}

\newcommand{\apjl}{Astrophys. J. Lett.}
\newcommand{\araa}{Annu. Rev. Astron. Astrophys.}

\newcommand{\grl}{Geophys. Res. Lett.}
\newcommand{\jgr}{J. Geophys. Res.}

\newcommand{\mnras}{Mon. Not. R. Astron. Soc.}
\newcommand{\ssr}{Space Sc. Rev.}
\newcommand{\pop}{Phys. Plas.}

\begin{document}
\title {Cosmic Ray Acceleration in Magnetic Reconnection Sites}

\author{Elisabete M. de Gouveia Dal Pino$^1$, Grzegorz Kowal$^2$, and Alex Lazarian$^3$
}

\affiliation{
$^1$ Instituto de Astronomia, Geof\'isica e Ci\^encias Atmosf\'ericas, Universidade de S\~{a}o Paulo -- IAG/USP,
Rua do Mat\~{a}o 1226, 05508-090 S\~{a}o Paulo, SP, Brasil.\\
$^2$ Escola de Artes, Ci\^encias e Humanidades, Universidade de S\~ao Paulo, Rua Arlindo Bettio 1000, CEP 03828-000, S\~ao Paulo , Brazil\\
$^3$ Department of Astronomy, University of Wisconsin, 475 North Charter Street, Madison, Wisconsin 53706, USA
\\
}

\begin{abstract}
Cosmic Ray (CR) acceleration still challenges the researchers. Fast  particles may be accelerated in astrophysical environments by a variety of processes. Acceleration in magnetic reconnection sites in particular, has lately attracted the attention of researchers not only for its potential importance  in the solar system context, but also in other astrophysical environments, like compact stellar sources, AGNs and GRBs,  and even in diffusive media like the ISM and the IGM, especially when the environment is magnetically dominated. In this talk we review this process and also present three-dimensional collisional MHD simulations with the injection of thousands of test particles, showing from the evolution of their energy spectrum that they can be efficiently accelerated  by reconnection through a first-order Fermi process within large scale magnetic current sheets (especially when local turbulence is present making reconnection fast and the acceleration layer thicker).
\end{abstract}

\vskip -1.35cm

\maketitle

\thispagestyle{fancy}

\setcounter{page}{1}

\bigskip

\section{CR acceleration: new challenges}

Recent years have witnessed new challenges on cosmic ray acceleration. On one side, the ultra-high-energy cosmic rays (UHECRs), which seem to be of extragalactic origin with source candidates ranging from compact stars to GRBs, AGNs or mergers of clusters of galaxies, still have their production mechanism(s) not fully understood \cite{kotera11}. On the other side,
very high energy observations of these sources with  Fermi and Swift satellites and
ground based gamma ray observatories indicate that relativistic particles are being accelerated up to $\sim$TeV energies in very compact regions \cite{sol13},  possibly magnetically dominated in some cases.

The mechanisms frequently discussed in the literature for accelerating energetic particles include varying magnetic fields in compact sources,
(e.g., \cite{degouveia00, degouveia01}),
stochastic processes in turbulent environments,
and acceleration behind shocks.
  The latter, in particular, is generally invoked as  dominating in several astrophysical environments (e.g., supernova remnants) and has been extensively discussed in the literature 
  (see the review by T. Bell  in these Proceedings).
An alternative, much less explored mechanism so far, involves particle acceleration within magnetic reconnection sites and in this highlight talk, we will discuss this mechanism and show that it may be also a very powerful mechanism for accelerating cosmic rays.

\subsection{Magnetic Reconnection}

Magnetic reconnection occurs when two magnetic fluxes of opposite polarity encounter each other (see Figure 1). In the presence of finite magnetic resistivity, the converging magnetic field lines annihilate at the discontinuity surface and a current sheet forms there.
Direct observations of reconnection indicate that at least in some circumstances, as in the solar corona and the Earth magnetotail,
it is $FAST$ with reconnection rates $V_{R}$ of the order of the Alfv\'en speed $v_A$. However, the standard  one-dimensional model of magnetic reconnection  proposed separately by Sweet (1957) and Parker (1958) \cite{sweet58, parker57}, the so-called Sweet-Parker (S-P) model (Figure 1, top panel),  using simple mass flux conservation arguments predicts a reconnection velocity given by $V_{R} \sim v_A (\Delta/L)   \sim v_A S^{-1/2} << 1$, where $S=L v_A/\eta$ is the Landquist number, with L being the large scale extension of the reconnection layer, $\Delta$ the thickness of the reconnection contact discontinuity,  $v_A = B/(4 \pi \rho)^{1/2}$ the Alfv\'en speed,  $B$ the reconnecting magnetic field,  $\rho$ the local density, and $\eta$ the Ohmic resistivity which is generally very small in the typically  high-conducting astrophysical plasmas,  resulting very large Landquist numbers. Thus, in contrast to the observations above,  the S-P model predicts a slow reconnection regime.

Petsheck (1964)~\citep{petschek64} proposed a way to solve this difficulty by assuming a two-dimensional geometry and making $\Delta \sim L$ by focussing the reconnection process
 into a single point, the X-point, rather than over the entire  large scale  L of the magnetic fluxes (see Figure 1, middle panel).  Such a configuration  indeed results a fast reconnection speed
$V_{R} \sim  \pi/4 (v_A ln S)$. However, it was later found to be unstable, rapidly collapsing to the S-P configuration in magnetohydrodynamical (MHD)  numerical simulations \cite{biskamp96}, unless a collisionless pair plasma with localized resistivity $\eta$ is considered
\cite{birn01, yamada10}.
In a collisionless two-fluid plasma, the particles' mean free path is of the order of the large scale dimension of the system ($L \sim \lambda_{mfp}$). Under such conditions, in an electron-ion plasma, for instance, the ion skin depth $\delta_{ion}$ (which can be viewed as the gyroradius of an ion moving at the
Alfv\'en speed, i.e. $\delta_{ion} = v_A /\omega_{ci}$, where $\omega_{ci}$ is the ion cyclotron frequency) is comparable to the S-P diffusion scale ($\Delta_{SP} =(L\eta/v_A)^{1/2}$). (In the case of an electron-positron pair plasma, a similar condition is valid for the electron skin depth.)  But at these scales, the Hall effect given by the $\vec{J}x\vec{B}$ term in Ohm's law is important and  able to sustain the Petschek-X-point configuration. In a collisional plasma, on the other hand,
  the  S-P  thickness is larger than
the microphysical length scales relevant to collisionless reconnection, i.e.,
$\delta_{ion} << \Delta_{SP}$ or $L > > \lambda_{e,mfp} (m_i/m_e)^{1/2}$ and the Hall effect is no longer dominant
\cite{shay98, shay04, yamada06}.

From the analysis above one could reach the conclusion that fast reconnection would be possible only in collisionless plasmas. Fortunately, this is not true and the reason is because
 nature provides an ubiquitous process that is turbulence and this makes reconnection fast in  collisional MHD fluids. This process was first described by Lazarian \& Vishniac (1999)  \citep{lazarian99} and
successfully tested with numerical simulations \citep{kowal09, kowal12}.
In a turbulent medium, the wandering of the magnetic field lines  allows for many simultaneous events of reconnection which make it fast (Figure 1, bottom panel). Indeed, one can demonstrate that $V_{R} \sim v_A (l/L)^{1/2} (v_l/v_A)^2$, where $v_l$ is the injection velocity of the turbulence and $l$ its injection scale. It is easy to see that for $l \sim L$ and $v_l \sim v_A$, $V_{R} \sim v_A$ and therefore, reconnection is fast! Besides, we see that  the presence of turbulence makes the reconnection layer thicker ($\Delta \sim l \sim L$) and intrinsically three-dimensional. Both  features will be very important for accelerating particles, as will see in Section 3. Later similar descriptions of fast reconnection in collisional MHD were also proposed by other authors (\cite{loureiro07,  shibata01}; see also \cite{uzdensky11} for a review).

\begin{figure}[!tb]
\includegraphics[width=\columnwidth]{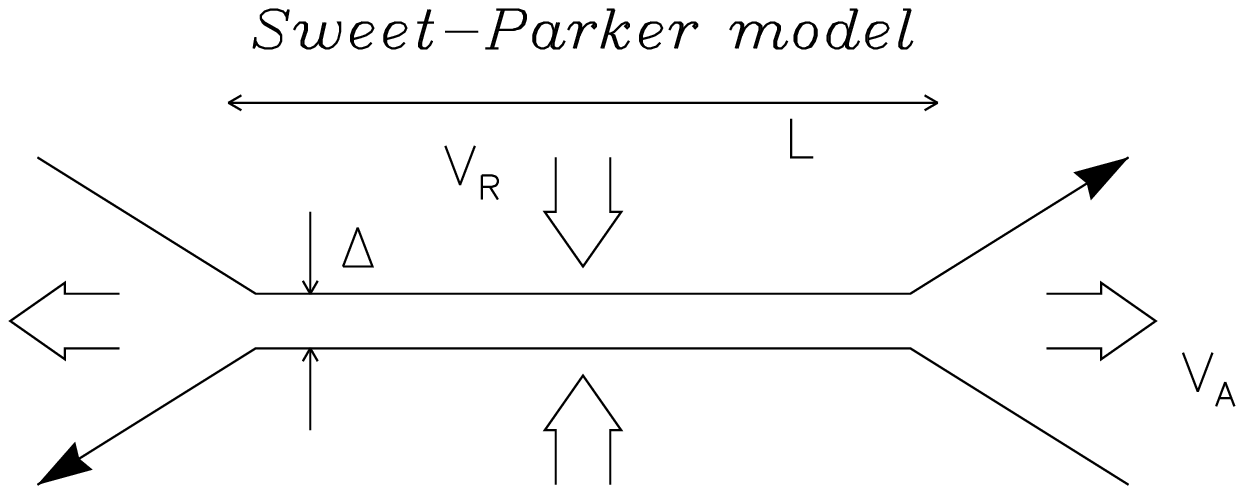}
\includegraphics[width=\columnwidth]{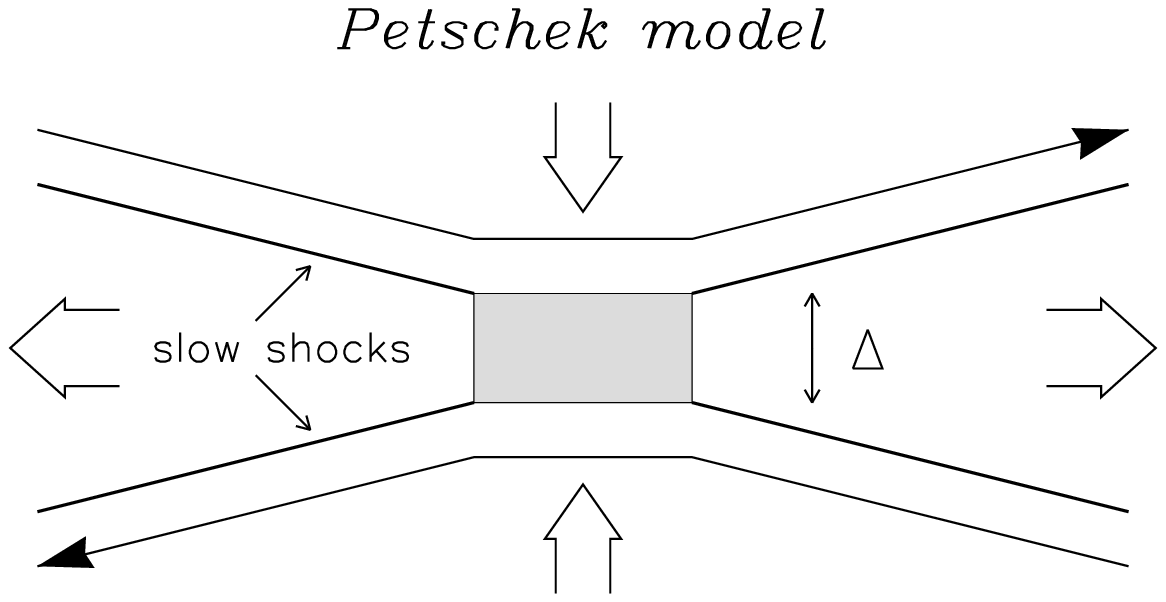}
\includegraphics[width=\columnwidth]{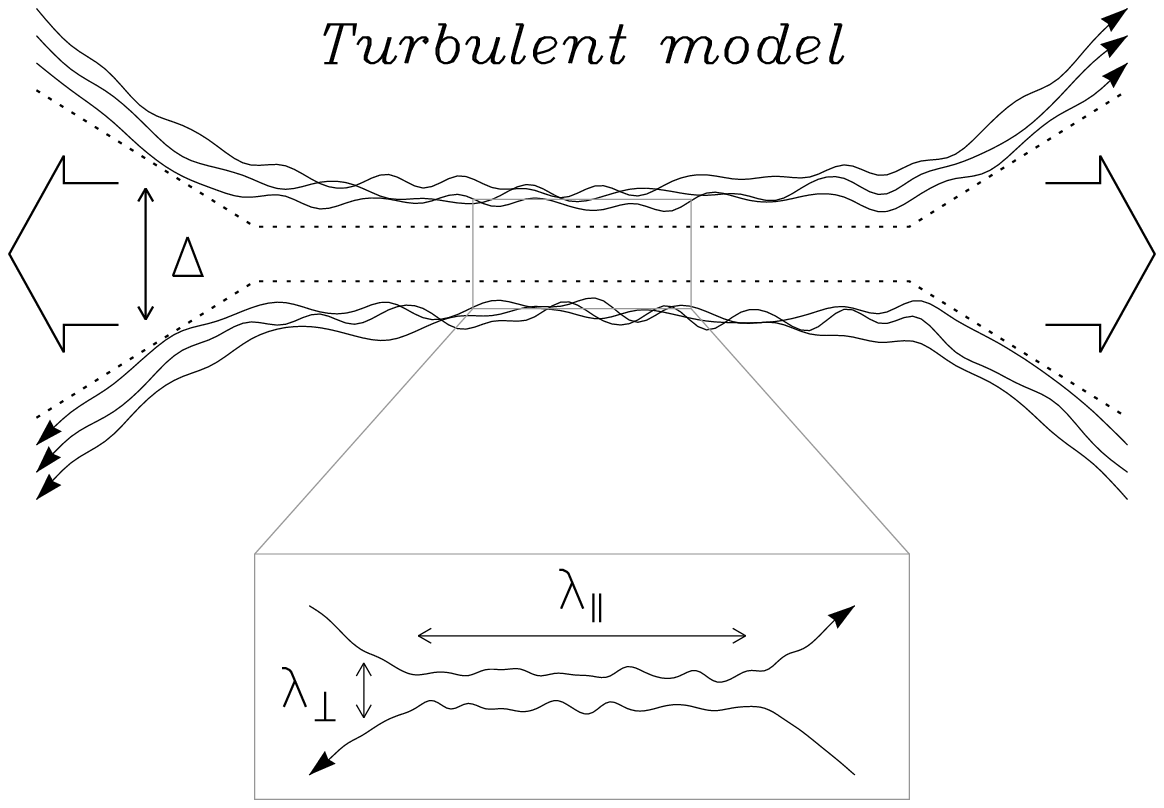}
\caption{\emph{\small
Schematic configurations of three different magnetic reconnection models. Top: Sweet-Parker model; middle: Petsheck model; and bottom: turbulent reconnection model. }
}
\label{fig1}
\end{figure}

\section{Magnetic Reconnection and First-Order Fermi Cosmic Ray Acceleration}

Fast reconnection breaks the magnetic field topology, releases  magnetic energy into the plasma in short time and is able to explain the bursty emission of, e.g.,  solar/stellar flares. In the later case, relativistic particle acceleration is always observed in connection with the flares. Thus the obvious question that one may pose is: can magnetic reconnection lead to direct efficient particle acceleration?
In the next paragraphs, we will discuss this issue.

Traditionally, particle acceleration in a reconnection site is thought as
due to the advective electric field created within the current sheet  along the z-axis (see top-left diagram in Figure 2). This reconnection electric field actually allows for a linear acceleration along the z direction and is given by $\epsilon_z=V_{R} B/c$ \cite{speiser65}.


In analogy to the shock acceleration mechanism where particles confined between the upstream (unshocked) and downstream (shocked) flows undergo a first-order Fermi acceleration (with a net energy increase after moving back and forth $<\Delta E/E> \sim v/c$, \cite{fermi49,bell78}),
de Gouveia Dal Pino \& Lazarian, in 2005 \cite{degouveia05}(henceforth GL05)  proposed that a similar mechanism would occur when particles are trapped between the two converging flux tubes moving to each other with $V_{R}$ in a magnetic reconnection current sheet (Figure 2). They showed that, as particles bounce back and forth  due to head-on collisions with magnetic fluctuations in the current sheet, their energy after a round trip increases by
 	$< \Delta E/E > \sim 8 V_{R}/3c$, which implies  a first-order Fermi process with an exponential energy growth after several round trips (see also \cite{lazarian11, degouveia13}). (Interestingly, Giannios (2010, \cite{giannios10}) assuming relativistic reconnection, obtained  an energy increase $ < \Delta E/E > \sim 4\beta_{R}/3 + \beta_{R}^2/2$, where $\beta=V_{R}/c$, which in the limit that $\beta_{R} < < 1$  recovers  the expression  obtained by GL05 for non relativistic reconnection.)

\begin{figure}[!tb]
\includegraphics[width=1.0\columnwidth]{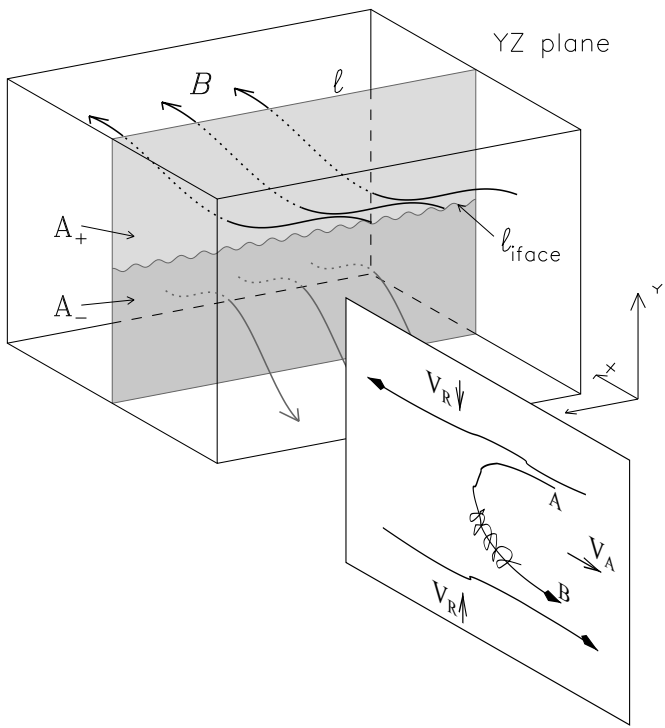}
\caption{\emph{\small
Top left: three-dimensional view of a magnetic reconnection sheet. Bottom right: detail of a particle being accelerated within the reconnection site. It spirals about a reconnected magnetic
field line and bounces back and forth  between points A and B. The reconnected
regions move towards each other with the reconnection velocity
$V_R$. Particles gain energy due to "collisions" with the magnetic irregularities within the two converging fluxes like in the first-order Fermi process in shock fronts (see GL05 \cite{degouveia05}).  Adapted from  \cite{kowal09, lazarian11}. }
}
\label{fig2}
\end{figure}

As remarked earlier, under
fast magnetic reconnection conditions, e.g., induced by turbulence
\cite{lazarian99}, $V_{R}$ can be of the order of the local Alfv\'en speed
$V_{A}$.
  In particular, at  the surroundings of relativistic sources  $V_{R} \simeq  v_A \simeq c$ and thus  the mechanism can be rather efficient \cite{degouveia05, degouveia10a, giannios10, lazarian05}.
   GL05 \cite{degouveia05} have also derived analytically  a power-law energy distribution for the accelerated particles
   ($N(E) \sim  E^{-5/2}$) and a corresponding electron synchrotron radio power-law spectrum  which is compatible with, e.g.,  the observed radio flares from galactic   black hole binary systems (or microquasars).

If one relaxes the assumption considered in the derivation  by GL05 that the particles escape from the acceleration zone with a similar  rate as in shock  acceleration, one obtains \cite{drury12}  a particle energy spectrum
 $N(E) \sim  E^{-(r+2)/(r-1)}$,  where $r = \rho_2/\rho_1$, and $\rho_1$ and $\rho_2$ are the densities in the inflow and outflow regions of the reconnection site, respectively.   Surprisingly, this gives the same power-law index as in shocks and, when  the compression ratio $r$ is large, $N(E) \sim E^{-1}$.
 We note however, that in the case of fast reconnection induced by turbulence \cite{lazarian99}, reconnection does not depend on the compression ratio.  On the other hand, one needs  to keep high anisotropy between the parallel and perpendicular components of the accelerated particle velocities in order to ensure efficient particle acceleration (see  \cite{degouveia13, lazarian11} and simulations below).

The considerations above also allow one to estimate
 the acceleration
time-scale due to reconnection.
The simplest way to evaluate this is
by setting the energy of the accelerated particle $E$ equal to $e (V_{R}/c)  B z$, where
z  is the distance travelled by the particle in the current sheet (normal to the large scale magnetic field direction) while being accelerated by the effective electric field $(V_{R}/c)  B$. This gives (e.g. \cite{speiser65, giannios10})
$t_{acc} \simeq  z/c \simeq \frac{E } {e V_{R} B}$.

Similarly, a simple way to estimate the maximum energy that a particle can reach  is by realizing that it  can no longer be confined within the reconnection region when its Larmor radius becomes larger than the thickness of the reconnection layer $l_{rec}$. This  implies that
$E_{max} \simeq  e c l_{rec}  B$.

\section{Is magnetic reconnection really a powerful process to accelerate particles?}

A way to probe the  analytical results above is through numerical simulations. So far, most of the numerical studies of particle acceleration by magnetic reconnection have been
performed  for two-dimensional (2D)  collisionless positron-electron plasmas by means of particle-in-cell (PIC) simulations    (e.g., \cite{drake06, drake12, zenitani01}). However, these apply to kinetic scales of only a few hundred  plasma inertial lengths  ($\sim 100 c/\omega_p$, where $\omega_p$ is the plasma frequency). The generally much
larger scales of the astrophysical systems (e.g., pulsars, AGNs, GRBs, etc) frequently require a collisional MHD description of reconnection.



Recent  studies in this direction undertaken  by \cite{degouveia10b, degouveia11, lazarian11, kowal11, kowal12}   modelled
 different domains of
magnetic reconnection  in two (2D) and three-dimensions (3D)  solving the isothermal MHD equations numerically  in a
 Godunov-type scheme.  10,000 test particles were injected in these MHD domains
with random initial positions and directions and with an initial thermal energy
distribution.  For each particle  the relativistic  equation of motion is solved
\begin{equation}
 \frac{d}{d t} \left( \gamma m \vec{u} \right) = q \left( \vec{E} + \vec{u} \times
 \vec{B} \right),
 \label{eq:ptrajectory}
\end{equation}
where $m$, $q$ and $\vec{u}$ are the particle mass, electric charge and velocity,
respectively, $\vec{E}$ and $\vec{B}$ are the electric and magnetic fields,
respectively, $\gamma \equiv \left( 1 - u^2 / c^2 \right)^{-1}$ is the Lorentz
factor, and $c$ is the speed of light.  The effective electric field $\vec{E}$ is taken
from the MHD simulations,
$ \vec{E} = - \vec{v} \times \vec{B} + \eta \vec{J}$,
where $\vec{v}$ is the plasma velocity, $\vec{J} \equiv \nabla \times \vec{B}$ is
the current density, and $\eta$ is the Ohmic resistivity coefficient. We note that the effect of the latter resistive term  on particle acceleration is
negligible \cite{kowal11}.

Figure 3 zooms in
 the details of the acceleration of a single test particle within
a resistive Sweet-Parker shaped current sheet \citep{kowal11}.
Before the particle reaches the current sheet discontinuity it is drifted by the
plasma inflow and the increasing gradient of B.
When it enters the discontinuity (the white part of the trajectory in
the upper panel), it bounces back and forth several times and gains energy (which
increases exponentially as shown in the detail of the lower panel of the Figure)
 due to head-on collisions with magnetic irregularities in the shrinking
flow, on both sides of the magnetic discontinuity, in a first order
Fermi process, as described in GL05 \cite{degouveia05}.  At the same time it drifts
along the magnetic lines which eventually allow it to escape from the
acceleration region.

\begin{figure}[!tb]
\includegraphics[width=\columnwidth]{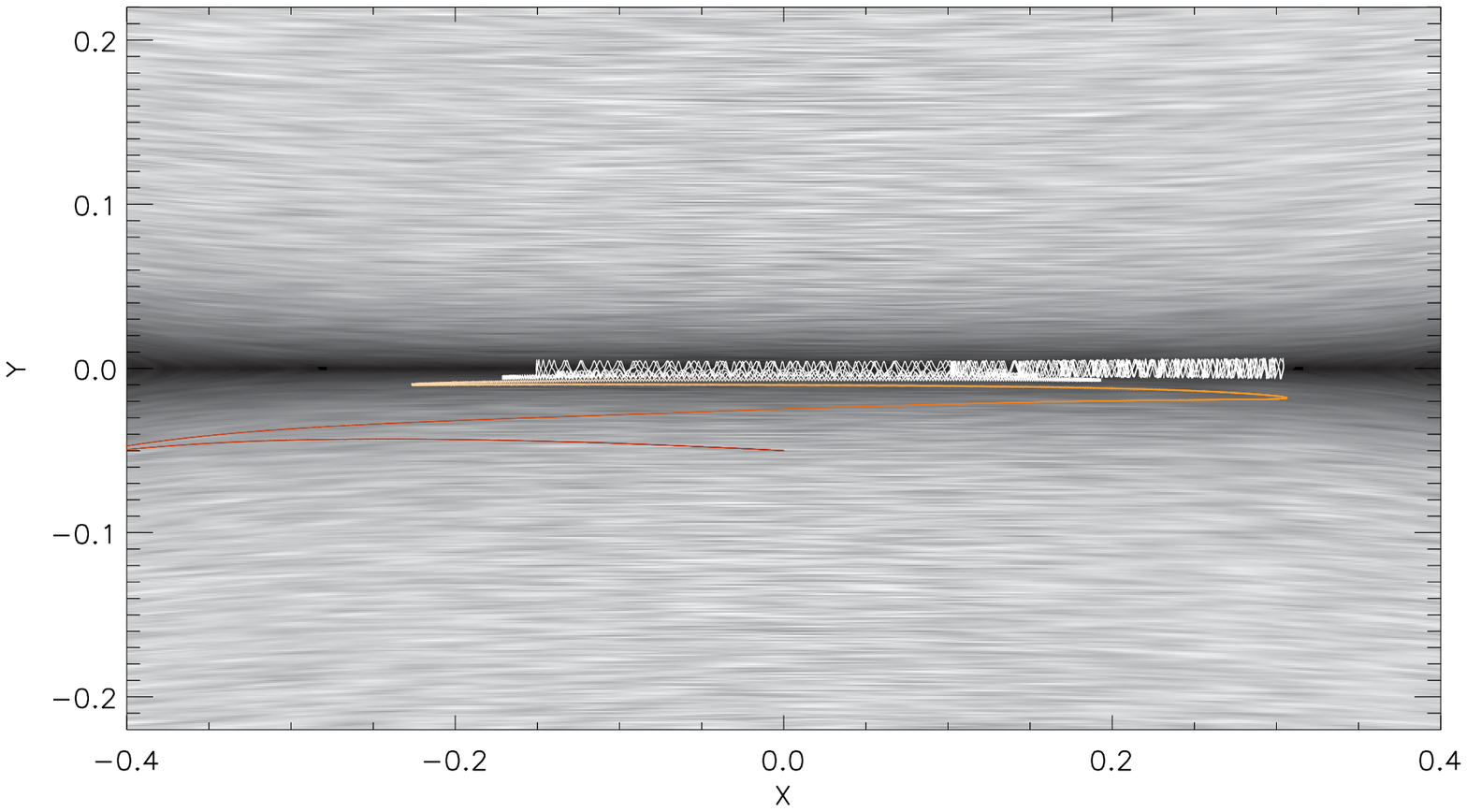}
\includegraphics[width=0.6\columnwidth]{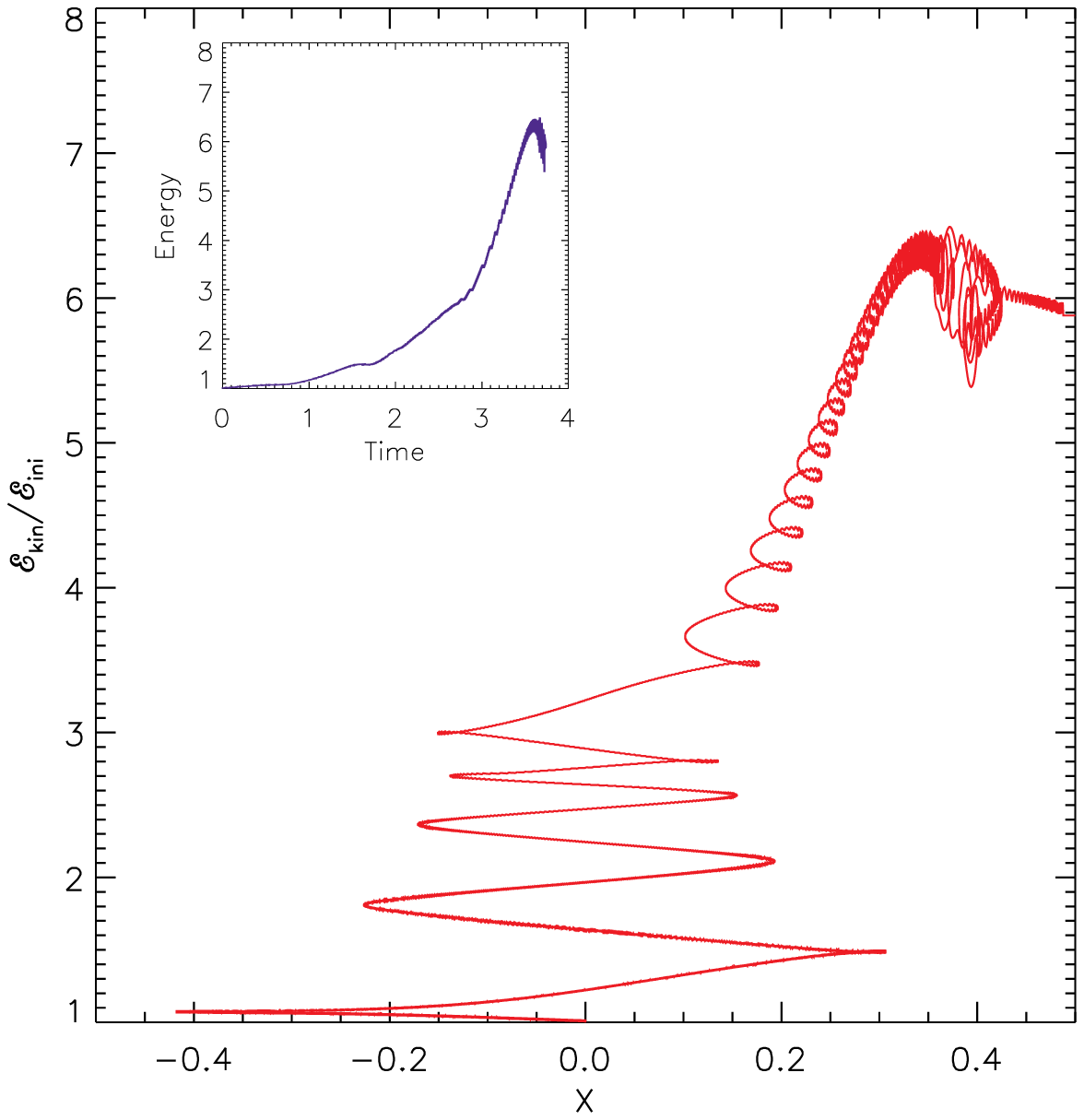}
\caption{\emph{\small
Numerical test of first-order Fermi acceleration within a  current sheet with a
Sweet-Parker configuration.  The top panel shows the trajectory of
a test proton approaching the discontinuity layer and bouncing back and force several times.  The color of the trajectory
corresponds to the particle energy (which increases from red to yellow and then
finally to white when the particle reaches the current sheet). The bottom panel
shows the evolution  of the particle energy which  increases exponentially when the particle reaches the current sheet.    In the model of
Sweet-Parker reconnection presented in this figure we used explicit large resistivity
coefficient $\eta = 10^{-3}$ in order to make reconnection fast.  The grid size in the model was set to $\Delta x =
1/1024$. From \cite{kowal11}. }
}
\label{fig3}
\end{figure}

Figure 4 presents an evolved 2D MHD configuration with  eight Harris
current sheets in a periodic box also computed by \cite{kowal11}.  Random weak velocity fluctuations were imposed to this
environment in order to enable spontaneous reconnection and the
development of 2D magnetic islands.
The merging of magnetic islands and the resulting stretching  or shrinking in some locations is evident.
The authors \cite{kowal11} find that
within the contracting magnetic islands and the
current sheets the injected 10,000 test particles accelerate predominantly through the first order
Fermi process, as previously described, while outside of the current sheets and islands the particles experience mostly drift acceleration due to magnetic
fields gradient. Their energy increases exponentially due to the Fermi process and the spectrum of the accelerated particles develops a high energy power law tail.
Similar results were found in 2D collisionless pair plasma PIC simulations 
\cite{drake06, drake10, drake12, zenitani01, cerutti13}.

\begin{figure}[!tb]
\includegraphics[width=\columnwidth]{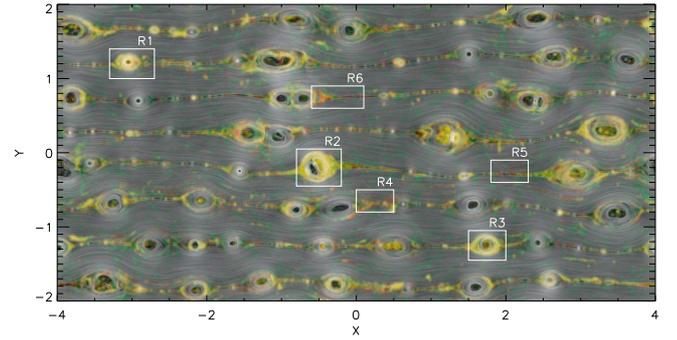}
\caption{\emph{\small
2D MHD configuration with eight current sheets and 10,000 test particles. The topology of the magnetic field is represented as a gray texture with
semi-transparent color maps representing locations where the parallel and
perpendicular particle velocity components are accelerated. There is no guide field
($B_z = 0.0$).  The red and green colors
correspond to regions where either parallel or perpendicular acceleration
occurs, respectively, while the yellow color shows locations where both types of
acceleration occur.  The parallel component increases in the contracting islands
and in the current sheets, while the perpendicular component increases
mostly in the regions between current sheets.  White boxes show regions that have been
more carefully analysed in \cite{kowal11}.  The simulation was performed with the
resolution 8192x4096.   The particles were injected  with
an initial thermal distribution with a temperature corresponding to the sound
speed of the MHD model (from \cite{kowal11}.) } }
\label{fig4}
\end{figure}

The results above confirm  that the
first-order Fermi acceleration process within shrinking islands is not restricted to  collisionless physics or kinetic effects as previously suggested   (e.g. \cite{drake06, drake10, drake12} and references therein).
This acceleration mechanism in reconnection sites works also in collisional plasmas, under the MHD approximation, as shown above and, in fact,
MHD codes present a way to study the physics of particle acceleration
numerically at large scales.
Acceleration in reconnection regions is thus a universal process which is not
constrained by the details of the plasma physics and can be also very efficient
in collisional gas.

These authors \cite{kowal11}  have also investigated  particle acceleration in 3D
MHD domains of reconnection and shown that its behaviour is  quite distinct from the acceleration in 2D domains.
In three-dimensions, the 2D  magnetic islands break into loops and the parallel component  of the particle velocities keeps accelerating in opened loops along the out-of-plane (z) direction. Acceleration in 3D reconnection is more like 2D reconnection with a mild guide field normal to the multiple current-sheet plane. In this case, as in 3D, the acceleration is not constrained to occur within merging islands in the plane only, but occurs also in open loops along the guide field \cite{kowal11}. However, if the guide field becomes too strong, then this will decrease the merge of islands and inhibit the overall acceleration process. This has been also found in studies of particle acceleration in collisionless flows (e.g., \cite{zenitani01,cerutti13}).

In the previous examples of particle acceleration in  collisional MHD, reconnection was made fast by means of enhanced numerical  resistivity. But, as remarked,   reconnection can be made  naturally fast when considering the universal presence of turbulence \cite{lazarian99}. Figure 5 shows a 3D MHD numerical simulation where \cite{kowal12} introduced
turbulence within a current sheet with a Sweet-Parker configuration and followed the trajectories of
10,000 protons injected in this domain.
As we see, an important
consequence of  the presence of  turbulent magnetic fields is the
formation of a thick volume filled with small scale magnetic fluctuations. This ensures a more efficient acceleration  than in the Sweet-Parker configuration
since the size of the acceleration zone and the number of scatterers is larger.
Figure 5 also  depicts the evolution of the kinetic energy of the particles.
 Particles are continually accelerated by encounters with several small and
intermediate scale current sheets randomly distributed in the thick shrinking volume and the
acceleration process is clearly  a first-order Fermi process, as in the previous cases.
We note that, after reaching the energy level $\sim 10^4$,
the particles accelerate at smaller rates, this is because their Larmor radii at this energy becomes comparable to the thickness of the acceleration region, and this is consistent with the maximum predicted energy evaluated in Section 2 for first-order Fermi acceleration. Further energy increase beyond this value is due
to a much slower drift acceleration (of the perpendicular component only) caused by the large scale magnetic fields gradients.

An inspection of the particle spectrum in the subplot of the bottom panel of  Figure 5 at t=5 c.u.  reveals already the formation of
a hard power law spectrum $N(E) \sim E^{-1}$  in the energy range $E/m_pc^2 \sim 10-10^3$, where $m_p$ is the proton mass.
This power law index is compatible with former results obtained from
2D collisionless PIC simulations considering merging islands  \cite{drake12}, or  X-type Petschek's configurations   (e.g., \cite{zenitani01}).

 Also from the simulations we can derive the acceleration
 time.  Preliminary  results from a number of  collisional MHD numerical simulations with  injection Alfv\'en velocities in the range $v_A/c \sim 1/1000- 1/5$
 \cite{kowal11, kowal12, delvalle13}, indicate that the acceleration time is nearly independent of the initial Alfv\'en (and   reconnection) speed)
and $t_{acc} \sim E^{0.4}$. This is generally longer than the estimated time in Section 2,  but becomes comparable to it as one approaches the maximum energy that the particles can reach in the acceleration zone \cite{delvalle13}.

It should be remarked  that the collisional MHD  simulations shown here focussed on proton acceleration. Although  applicable to electrons too,  the numerical integration of the electron trajectories is much longer. Nevertheless, such tests are also needed.

Those authors \cite{kowal12}  have  also tested the
acceleration of particles in  $pure$ MHD turbulence, where particles suffer
collisions both with approaching and receding magnetic irregularities. The
acceleration rate is smaller in this case and suggests that the dominant process is a
second-order Fermi.
We note that other recent studies have also explored test particle acceleration both in turbulent and in resistive MHD domains (see \cite{dmitruk03, onofri06, gordovskyy10, gordovskyy11}), 
but did not explore  the nature of the mechanism  accelerating  the particles.

\begin{figure}[!tb]
\begin{center}
\includegraphics[width=\columnwidth]{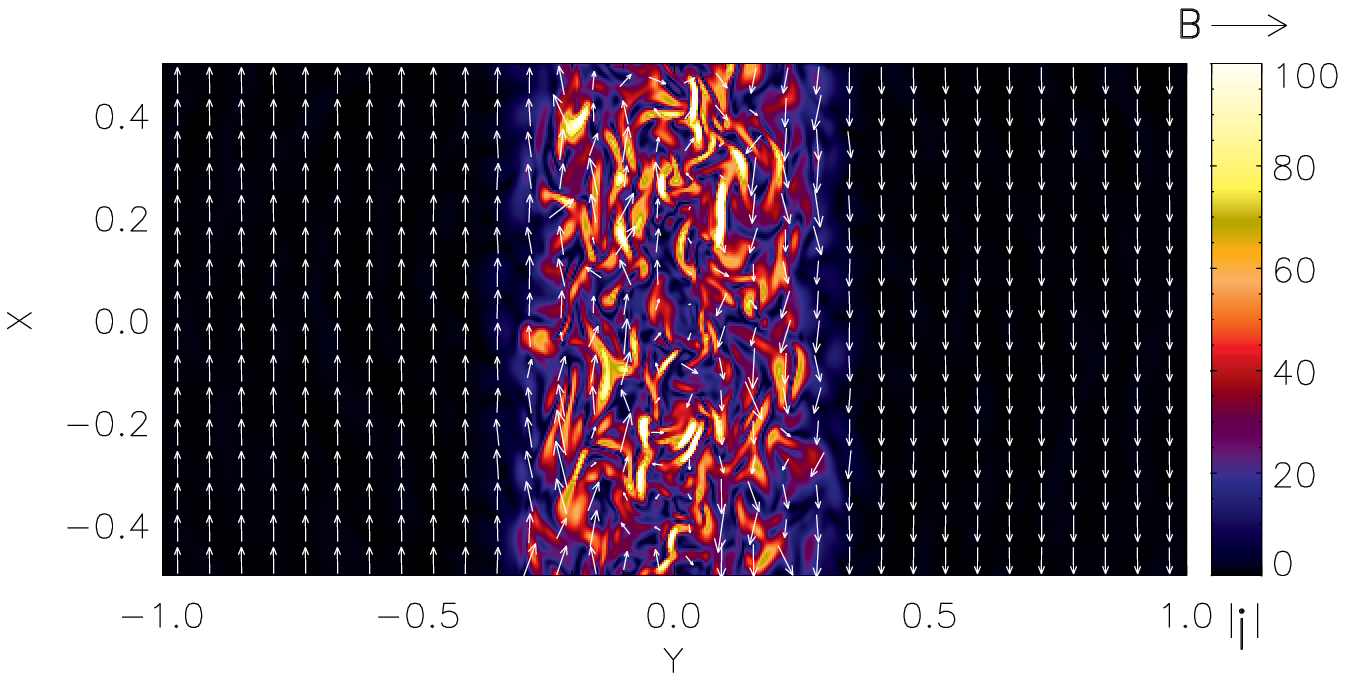}
\includegraphics[width=\columnwidth]{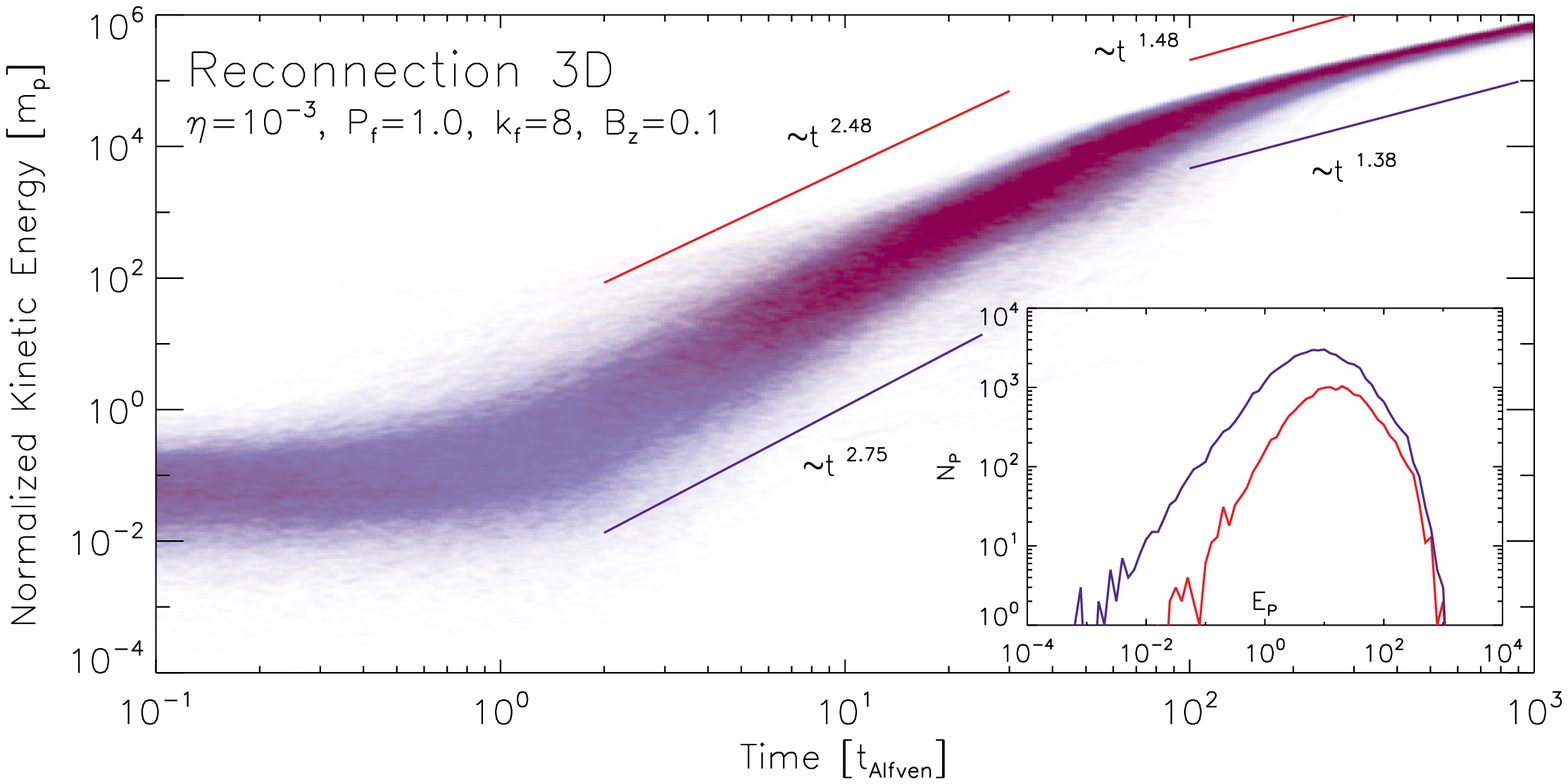}
\end{center}
\caption{\emph{\small
 Bottom panel: particle kinetic energy distributions for 10,000
protons injected in a 3D current sheet with turbulence which makes reconnection fast.  The colors
indicate which velocity component is accelerated (red or blue for parallel or
perpendicular, respectively).  The energy is normalized by the rest proton energy.
The subplot shows the particle energy distributions at $t=5.0$.  Top panel:
 XY cuts through the domain at $Z=0$ of the absolute value of
current density $|\vec{J}|$ overlapped with the magnetic vectors. It was employed $B_{0z} = 0.1$, $\eta=10^{-3}$, and a
resolution 256x512x256 (from \cite{kowal12}.) } }
\label{fig5}
\end{figure}

\section{Particle acceleration in relativistic reconnection sites}
In this review,  we discussed mostly  Fermi  acceleration considering non-relativistic reconnection environments, that is, generally assuming $V_{R}$   smaller than the light speed.  However, in systems like, e.g, the  surroundings of black holes and pulsars, $v_A \sim  c$  and then,  since  in fast reconnection  $V_R$ must  approach  $v_A$, reconnection itself may become relativistic in such domains.
There has been some advancement in relativistic reconnection studies too. The
theoretical grounds have been established by a number of authors (\cite{lovelace92, blackman94, lyubarsky05, lyutikov03, jaroschek04, hesse07, zenitani08, zenitani09, komissarov07, coroniti90, lyubarsky01}; see also \cite{uzdensky11, degouveia13} and references therein for  reviews).
Essentially, it has been found that  Sweet-Parker reconnection in the relativistic regime is slow, as in the non-relativistic regime,  while  2D X-point Petschek's  reconnection  predicts a fast  rate as in the non-relativistic regime \cite{lyubarsky05}.  The numerical advances  in relativistic reconnection have been performed so far only for 2D collisionless X-point  Petschek's  configurations  by means of PIC simulations of pair plasmas, and have confirmed the results of the  analytical theory.  In such relativistic collisionless electron-positron pair plasmas, the investigation of relativistic particle acceleration  is almost  straightforward. Studies  e.g.,  by  \citep{zenitani01} have revealed results which are compatible with those of acceleration in non-relativistic reconnection.

However, studies of particle acceleration in the collisional relativistic MHD  regime (RMHD) are still in their childhood.  Preliminary results of numerical simulations of $in$ $situ$ particle acceleration in relativistic jets (see Figure 8 of \cite{degouveia13}) indicate that both, shock acceleration and acceleration by magnetic reconnection are competitive mechanisms.

 These results are encouraging and may have rather important consequences  on particle acceleration and high energy emission processes  in  compact sources (see below).

\section{A few applications}

As remarked, magnetic reconnection is very frequent and therefore, it should be expected to induce acceleration of particles in a wide
range of galactic and extragalactic environments.  Originally discussed
predominantly in the context of solar flares \cite{drake06,
drake09, gordovskyy10, gordovskyy11, zharkova11} and earth magnetotail \cite{lazarian09,
drake10,lazarian10}, it has been gaining importance beyond the solar system, in more
extreme astrophysical environments and sources, such as in the production of  ultra
high energy cosmic rays  \cite{giannios10,
kotera11}, in particle
acceleration in jet-accretion disk systems \cite{degouveia05, degouveia10a, degouveia10b,
giannios10, delvalle11}, and in the general framework of compact sources, as AGNs and GRBs
\cite{lazarian03, zenitani01, zenitani09, degouveia10b, giannios10, zhang11, uzdensky11,
degouveia11}, and even in pulsar nebulae, like Crab \cite{cerutti13}.

For instance,  assuming multiple magnetic field reversals along the jet as in Figure 3 (imprinted at the jet launching region near the light cylinder),   \cite{giannios10} estimated that AGN and GRB jets would be able to produce UHECRs through first-order Fermi acceleration by magnetic reconnection (GL05, \cite{degouveia05}).
The multiple reversals can easily produce a power law spectrum of relativistic particles, as we have seen, e.g. in \cite{kowal11}, but this model still requires numerical testing using   specific conditions of relativistic jets in order to see whether such high energies can be really achieved.


Another potential site for particle acceleration by reconnection has been explored by \cite{degouveia05, degouveia10a, degouveia10b, kadowaki13}  at the corona of  accretion disks around compact sources (like AGNs and  galactic BHs). They computed the magnetic power arising from fast reconnection in the inner disk region and found a correlation with the mass of the sources over  $10^9$  orders of magnitude in mass. Plotting in the same diagram the Synchrotron radio power observed for more than 100 galactic BHs  and different classes of AGNs, they found that the predicted magnetic reconnection power is more than sufficient to explain particle acceleration and the observed radio and gamma outbursts from both galactic BHs and low luminous AGNs, but not those from high luminous AGNs. This result might offer a physical interpretation for the so called $Fundamental$ $Plane$ \cite{merloni03} which correlates the emission of galactic BHs and low luminous AGNs. They also concluded  that the lack of correlation with high luminous AGNs (e.g., BL Lacs for which the jet is oriented towards the line of sight) is due to the fact that in such cases the nuclear emission is screened by the high density and radiation fields around the nucleus and what is observed is the emission from the jet further out is such cases. This result is also compatible with recent findings of \cite{nemmen13} (see also \cite{kadowaki13}).


Another example explores the potential effects of acceleration by reconnection in the Crab nebula. Puzzling Synchrotron flares with energy larger than the classical ideal MHD limt $E > E_{max} \sim   9mc2/4sF \sim$ 160  MeV have been observed in this source and attributed to
  randomly oriented relativistic “mini-jets”
driven by reconnection (see \cite{clausen-brown12}). These authors evaluated the acceleration of the particles due to reconnection in such mini-jets  and concluded that this   would be able to reproduce the observed spectral energy distribution. It is interesting that this model presents features which resemble the fast reconnection model of acceleration mediated by turbulence as  in Figure 4.

\section{Forthcoming studies}

Cosmic ray acceleration investigation in magnetic reconnection sites is still in its youth, particularly in collisional MHD and  relativistic regimes. Current studies in this regard include   relativistic MHD fast reconnection involving  turbulence (e.g. \cite{degouveia13}) and
relativistic reconnection of electron-ion, high energy density, radiative plasmas  (e.g. \cite{uzdenski11}). These are  fundamental issues  specially for  modelling flares and variability in the spectrum of  compact sources, like AGNs, GRBs, microquasars, etc.

 Also,   particle acceleration in  diffuse domains of reconnection, particularly in pure turbulent regions like the ISM and IGM, deserves attention as these particles
may be available, e.g.,  as  seed populations for further acceleration in shocks or magnetic reconnection regions in embedded sources.

Forthcoming studies will also require the inclusion of the relevant loss
mechanisms of the accelerated particles, in order to assess the importance of the acceleration by reconnection
 in comparison to other processes (e.g.,  shock acceleration)
and to reproduce the observed spectra and light curves of the sources (see e.g., \cite{cerutti13,
khiali13}).

In conclusion,
 magnetic reconnection is now recognized as an essential process not only in the solar system, but also beyond it, in a large number of astrophysical sources, including   turbulent environments which in turn, are  ubiquitous.  In this situation the acceleration of cosmic rays by reconnection may play a vital role and deserves extensive investigation, specially to help in the interpretation of current  high energy  observations and in making  predictions for  upcoming new generation of instruments, like the Cherenkov Telescope Array (\cite{cta11, cta13, sol13, barres13}).

\subsection{Acknowledgements}
 E.M.G.D.P. acknowledges  partial support from the Brazilian agencies FAPESP (grant no. 2006/50654-3 and CNPq  (grant no. 300083/94-7) and G.K. also acknowledges support from FAPESP (grant 2009/50053-8).  Part of the simulations presented here have been carried out in LAI (Astrophysical Laboratory of Informatics at IAG-USP).

\end{document}